# Wave propagation through alumina-porous alumina laminates

Pathikumar Sellappan[1], Erheng Wang[2], Christian J. Espinoza Santos[1], Tommy On[2], John Lambros[2] and Waltraud M. Kriven[1*]

[1]Department of Materials Science and Engineering, University of Illinois at Urbana-Champaign, Urbana, IL, USA

[2]Department of Aerospace Engineering, University of Illinois at Urbana-Champaign, Urbana, IL, USA

* Corresponding author:

Professor Waltraud M. Kriven

E-mail: kriven@illinois.edu

**Abstract**

A Brazilian disk geometry of an alumina layered composite with alternating dense and porous layers was dynamically loaded using a Split-Hopkinson Pressure Bar (SHPB) apparatus under compression. High-speed imaging and transmitted force measurements were used to gain an insight into stress wave propagation and mitigation through such a layered system. Uniformly distributed porosities of 20 and 50 vol % were introduced into the interlayers by the addition of fine graphite particles which volatilized during heat treatment. Brazilian disk samples were cut from the cylinders which were drilled out of the sintered laminated sample. The disks were subjected to dynamic impact loading in perpendicular and parallel orientations to the layers in order to investigate the influence of the direction of impact. The dynamic failure process of the layered ceramic consisted of the initiation and propagation of the cracks mainly along the interphases of the layers. Upon impact, the impact energy was dissipated through fracture in parallel orientation (0°) but transmitted in perpendicular (90°) orientations. The high degree of correlation between the transmitted force, microstructure and orientation in which the layered systems were impacted is discussed.

1. Introduction

Even though ceramics exhibit unpredictable failure, various components in the form of monolithic tiles, coatings and fibers have been extensively employed in structural applications, especially as armor materials[1–3]. Due to their attractive mechanical strength properties, both oxide (mainly alumina and zirconia-based)[4] and non-oxide (carbide, nitride, boride, etc.,)[5–8] based structural ceramics have attracted considerable attention for use under impact loading. In armor usage, when a hard projectile impacts a ceramic object, the impact area might be fractured, pulverized, and/or ejected, depending on the dynamic impact conditions[1,9]. Fracture as well as fragmentation of the impact plates are effective ways to dissipate the impact energy which consequently protects the backing surface[1]. The extension of time of impact and redistribution of the impact load over a wider area on the supportive backing structure helps to reduce the stress concentration during the dynamic failure process[1]. The ability to guide and deflect cracks enhances the energy dissipation and this can be achieved using interphases that are weaker than stiff plate materials[2,9]. Furthermore, the complex dynamic compressive behavior of brittle solids such as ceramics, rocks and concretes has been studied extensively using a Split Hopkinson Pressure Bar (SHPB) arrangement[3,10,11]. The SHPB, which was originally developed by Kolsky[12], has been modified to determine the dynamic deformation behavior of materials under controlled strain rates ($10^2$/s–$10^4$/s). In reality, such extreme loading conditions are indeed changing the way that brittle materials usually fail on an atomic scale[13].

There are several research efforts that have been made to understand the role of different parameters (strain rate, microstructure and environmental effect, etc.) on the fracture and failure of ceramics under dynamic loading. However, in comparison with our understanding of the microstructure-property relation under static loading, the dynamic mechanical response of structural ceramics is still a young and topical engineering problem. Simultaneously possessing all the basic requirements such as low density, high strength and high toughness to design better ballistic protection is not possible using the available engineering materials alone. Therefore, concepts of layered structures have been successfully implemented for armor applications[9,14,15]. Ceramic layered systems have attracted wide attention due to the crack deflection capability in the weak interlayers which has been shown to be effective in improving toughness of components[16–20]. In these layered ceramic systems, toughness improvement is the result of crack deflection which mainly depends on the fracture energy absorbed in the interphase of the laminates[20]. Current literature on the high-strain rate deformation behavior of laminated systems is either limited to biological and metallic systems or compressive stress-strain analyses in the case of ceramic laminates[21–23]. However, layered structures of such brittle constituents are not only present in man-made materials but also in naturally available materials like bone, nacre and the conch shell[24,25]. Understanding the deformation behavior of naturally available, brittle layered systems can help us to design better armor components[24,26]. For example, in biological systems such as *Nacre* and *Strombus Gigas*, mechanical behavior depends on the orientation in which a sample was tested, due to their complex layered microstructures[25]. In the past, some attempts have been made to mimic naturally available systems to process layered ceramics at the microstructural scale[26,27], but testing under various orientations, and under dynamic loading has not been performed until now, in part due to severe difficulties associated with fabrication of appropriate bulk samples.

Investigating stress-wave mitigation in layered systems based on the orientation, strain rate, varying the weaker layer thickness ratio, and density can help to design better materials for ballistic protection[1]. Determining constitutive and failure models of dynamic impact studies will help to develop numerical simulations which will give critical information on armor performance[14,28,29]. In order to obtain reliable and reproducible predictive models, materials data under high-strain rate or high pressure or a combination of both are highly desired.

In the present study, we successfully implemented key processing features to fabricate a model system, in the shape of a Brazilian disk, consisting only of alumina, with alternative dense (stronger) and porous (weaker) layers in its microstructure. Our approach was not only selected to investigate the stress-wave mitigation behavior in a specific and desired direction in a controlled way, but also to visualize the dynamic failure and fracture processes easily. The alumina/porous alumina combination was chosen based on previous studies on the importance of a chemically compatible interphase so as to avoid the accumulation of internal residual stresses[30–32]. The best way to achieve such a system is to fabricate laminates with porous interlayers of a material of composition which is the same as that of the dense material[20,30]. In this study, we used fine graphite particles as pore formers to introduce uniformly distributed, fine, architectured porosity into the porous interlayers. Furthermore, the present investigation focuses on the dynamic force response with respect to the orientation of the layers, rather than on the compressive stress-strain behavior. A modified SHPB apparatus with a momentum trap was employed to investigate wave propagation and the load transmission results are discussed in correlation with high-speed imaging results.

## 2. Experimental Procedure
### 2.1 Laminate Fabrication

Dense and porous alumina green tapes were produced using commercially available alumina (A16SG, Almatis, Leetsdale, PA, USA) and graphite powders (Aldrich Chemical Company, St. Louis, MO, USA) as precursor materials. The fine alumina powders were 99.8% chemically pure with an average particle size of ($D_{50}$) 0.45 $\mu$m and a surface area of 8.5 $m^2$/g, and had a density of 3.98 ± 0.01 g/$cm^3$. Graphite particles with appropriate amounts of 0, 20, and 50 vol %, were introduced into the slurry which contained A16SG alumina particles. The graphite particles used as pore formers had a 1-2 $\mu$m particle size and a 1.9 g/$cm^3$ density. A formulation supplied by Polymer Innovation, Inc® (Vista, California, USA) was used to prepare the slurries and it contained (i) acrylic binder (WB4101) which consisted of defoamer, plasticizer, and resin to produce basic tapes, (ii) a non-silicone mild defoamer (DF002), and (iii) a high pH plasticizer (Pl002). This polymer was a combination of strong dispersing molecules and partly strong binder molecules to help in milling the slurry without excessive foam or destabilization.

Powders, de-ionized water and appropriate amounts of binder were added at the first stage to obtain very low viscosity slurries which later resulted in a stable suspension. The slurry compositions used to prepare both graphite-containing and graphite-free tapes are given in Table 1. The ingredients were mixed by ball milling (at 92 rpm for 16 h) using yttria stabilized zirconia cylinders as milling media. The remaining ingredients were then added and ball milling was continued for 4 more hours. After milling, the resulting slurries were drained and sieved through a mesh, followed by de-airing using a vacuum desiccator and left in a fume-hood for five minutes to remove any existing air bubbles. The slurries were then tape cast on to Mylar sheets using a hand-held doctor blade having an initial thickness setting of 250 μm, which was then reduced to ~ 125 μm thickness after drying in air.

After drying, the tapes were cut into 38 mm squares, stacked, and pressed together in a custom designed die at 72 °C, under a pressure of 10 MPa for 10 minutes. The pressure was increased to 20 MPa and held for 20 minutes to fabricate samples of ~ 37 mm thickness, as shown in Fig. 1(a). To investigate the role of porous layers under dynamic impact, samples were made by alternatively stacking graphite-free (dense alumina) and graphite-containing (porous alumina) tapes. In all the cases, 5:1 ratios of dense alumina to porous alumina layers (20 and 50 vol %) were maintained. For comparison, monolithic dense alumina was also fabricated using the same technique but without graphite addition. These are hereafter referred to as dense alumina, 20 vol % and 50 vol %, respectively. Binder and graphite burnout were carried out based on thermogravimetric analysis (TGA, Netzsch STA 409 CD, Selb, Germany) performed in air from room temperature to 1300 °C. The laminates prepared in this study were slowly heated to 900 °C at a heating rate of 0.5 °C min$^{-1}$ and held for 1 h to remove the pore formers (graphite particles) and binders. Then, the laminates were heated to 1550 °C at a heating rate of 1 °C min$^{-1}$ and held for 5 h in air.

Cylinders of 25.4 mm diameter were then drilled from the sintered specimens as shown in Fig. 1(b). Diamond-impregnated core drills (Wale Apparatus, Hellertown, PA, USA) were employed to drill the specimens by aligning the loading axis parallel to the layer planes, using manual drilling equipment (Lunzer, Industrial Diamond Inc, New York, NY, USA). Subsequently, Brazilian disk specimen of ~3 mm thickness were sliced using a low speed, diamond-tipped saw from the cylinders which were drilled out of the sintered laminates. The disks were ground in a "Buehler Ecomet III" polishing apparatus, using diamond polishing disks and polishing pads (Buehler) down to a 3 μm finish to obtain smooth and parallel surfaces (Fig. 1(b)).

## 2.2 Dynamic mechanical loading

A schematic of the SHPB apparatus used to evaluate the dynamic responses of the dense alumina disks as well as the laminated disks is shown in Fig. 2. The diameter of the incident, transmitted and striker bars was 12.7 mm and all bars were made of maraging steel C350 having a density of 8.10 g/cm$^3$, Young's modulus of 200 GPa and an elastic bar wave speed of about 5000 m/s. The laminated Brazilian disks were sandwiched between the incident and transmitted bar at 0° (laminate axis was parallel to the incident and transmission bar) and 90° (perpendicular direction) orientations. In all cases, a lead pulse shaper (< 1 mm) was placed between the impacting striker and the incident bar in order to control the incident pulse rise time and profile so as to avoid sample premature failure. In addition, the momentum trapping technique was adopted to prevent multiple loadings of the specimen. Once the gas gun was triggered, the striker bar was launched towards the incident bar and an elastic compressive wave was generated within the incident bar, upon impact. At the interface between the sample and the bars, the wave was partially transmitted through the sample and partially reflected back.

The incident ($\varepsilon_i$), reflected ($\varepsilon_r$) and transmitted ($\varepsilon_t$) strain profiles were obtained from measured signals on the bars, and the corresponding input and output forces of the Brazilian disk specimen were calculated using the following equations[23,33],

$$F_{input} = -(\varepsilon_i(\tau) - \varepsilon_r(\tau))AE \qquad (1)$$

$$F_{output} = -\varepsilon_t(\tau)AE \qquad (2)$$

where $\tau$ is the time (with signals being time shifted to be time coincident with the arrival of each wave at the incident bar/specimen interface), and $A$ and $E$ represent the cross-sectional area and Young's modulus of the bar, respectively. Typically when the SHPB is used to determine uniaxial compressive response of bulk material strain rates in the range $10^2$ s$^{-1}$ to $10^4$ s$^{-1}$ can be achieved[33]. However since the strain and strain rate fields in the Brazil disk configuration are highly inhomogeneous (concentrated near the contact points) and average strain rate cannot be easily defined. We define the diametrical deformation rate (Ṡ) to quantify the rate at which the diametrical dimension of the Brazilian disk between the two contacts points changes. It can be measured using the following equation

$$\dot{S} = \frac{S_{max}}{t_{crack}} = \frac{-\int_0^{t_{crack}} c\left(\varepsilon_i(\tau) + \varepsilon_r(\tau) - \varepsilon_t(\tau)\right) d\tau}{t_{crack}} \quad (3)$$

where Ṡ is the diametrical deformation rate, C is speed of sound in the bar (approx.. 5000 m/s), $S_{max}$ is the maximum deformation at time $t_{crack}$ and $t_{crack}$ is the time when the crack initiates. If the specimen is intact, $t_{crack}$ is the total loading time.

Dynamic compression studies were performed on a minimum of four specimens in each category (dense alumina, 20 and 50 vol % laminates). Deformation and dynamic failure processes of the specimen were recorded using a high speed digital camera (FASTCAM, SA5 Model 775k-M3, Tech Imaging, Salem, MA) at a frame rate of 100,000 fps and an exposure time of 1 μS. The captured images were then synchronized manually based on the specimen movement with the input force signals to analyze the failure and fracture process during the dynamic impact event.

### 2.3 Material Characterization

The Archimedes method (ASTM C373) was employed to measure the average bulk densities of the specimens. Optical (Leica MZ6, Wetzlar, Germany), and scanning electron microscopic (SEM, JEOL 6060LV, Tokyo, Japan) analyses were performed on the polished surfaces of the specimens, as well as the dynamically fractured specimens in order to carry out the microstructural investigation. Both sides of the disks were marked in five locations, to trace the crack propagation by optical microscopy after the experiment.

## 3. Results and Discussion
### 3.1 Processing of Laminates

The green tapes obtained by an aqueous-based formulation were of uniform thickness, had smooth surfaces, and were free of visible pinholes after being dried in air.

The TGA analyses performed at a heating rate of 5 °C/min on alumina with 0, 20 and 50 vol % graphite particle contents is shown in Fig. 3. The tapes underwent three stages of drying viz.: loss of water at an early stage (< 200 °C), organic removal (~ 400 - 600 °C) and graphite removal (> 600 °C). The TGA results clearly revealed that the polymer content in the tapes could be removed completely at around 500 °C in the case of graphite-free alumina tapes. However, weight loss continued until 900 °C when the graphite content increased to 50 vol % in the slurry. In all cases, the tapes completely lost all of the volatile materials before 900 °C and there was no weight loss observed after that. Furthermore, the optimum sintering conditions (1550 °C, 5h) yielded > 98% dense alumina in the case of graphite-free alumina (Table 2). In the present case, the fine particle size of the starting material, high surface area and good packing due to the processing resulted in > 98% dense material compared to the theoretical density of phase pure α-alumina (3.981 g/cm$^3$). However, due to the oversize of the sample, it was decided to heat and cool down at a rate of 1 °C/min to avoid any thermal shock. Hence, the same thermal cycles were followed for all the samples. The ceramic laminates were able to survive the severe drilling which resulted in a uniform cylinder of ~ 25.4 mm diameter and length of ~ 30 mm.

SEM micrographs of cross-sectioned, sintered alumina made from graphite-free tapes are shown in Fig. 4(a) and (b). The surface analysis of alumina showed that the sintered samples had a homogeneous microstructure with uniformly distributed submicron level porosity (pore sizes were <1 µm), and the tapes were well fused together at the interfaces without any microscopically observable defects. As can be seen in Fig. 4(c) and 4(d), which shows the magnified microstructure of the thermally etched surfaces, the grains in the pore-free region of alumina had grown to more than 13 µm in size with an average grain size of 2.6 µm (within the range of 0.5 to 13.2 µm). Preferential, exaggerated grain growth is a common feature in alumina when it is sintered without any dopant or sintering additives[34]. The average pore size measured by SEM analysis was 0.65 µm (having a range of 0.2 to 1.6 µm).

Microstructural observations of the cross-sectioned alumina/porous alumina layered systems are shown in Fig. 5. It can be seen that the layers were very well integrated and free of any initial delamination between the dense and porous layers. This helped us to scale up the dense or porous layers as well as the sample thickness to desired sizes without compromising the dimensional requirement for mechanical evaluation. The microstructures from Fig. 5(a) to (d) show that the porosity level varied in the porous layers as the graphite particle content increased from 20 to 50 vol %. This shows that the dense alumina layers were fully densified during the sintering process and the addition of graphite produced porosity throughout the porous layer. In general, the pores were fine (less than or equal to 20 μm) with random shapes and were uniformly distributed in the alumina matrix. Microstructural analysis of the porous alumina layers revealed that addition of graphite particles into alumina to increase the porosity resulted in not only fine grains in the porous layers, but also refined microstructure and more homogeneous grains as well as a pore distribution (Table 2 and Fig. 5). The average grain and pore sizes measured by SEM for the porous layers prepared with the 20 vol % graphite was 1.9 μm and 4.6 μm, respectively. However, when the starting graphite particle content increased to 50 vol %, the grain size in the porous layers was reduced to 1.3 μm and the pore size was 3.3 μm with a more homogeneous distribution (Table 2).

When alumina particles mixed with graphite particles underwent heat treatment, the particles were separated after removal of polymers and pore-formers. Hence, the regular diffusion process of sintering was hindered which resulted in porosity. The particles underwent rearrangement as well as some local densification. The slow heating and cooling rates could have provided enough time to complete the sintering process resulting in a more homogeneous microstructure. Earlier studies using porous systems showed that pores with lower coordination numbers tend to sinter or otherwise they remained stable[31,35]. Also, a previous study on fabricating porous alumina samples using the same precursors but a different route and heat treatment produced an increase in the apparent porosity by increasing the pore-formers[36]. Unlike some other studies which used pore-formers such as PMMA or starch and also different processing routes resulting in macropores, the present study resulted only in micropores. Since the microstructure of dense alumina layers remained the same in the lamellar structure, the changes in the density values could be correlated to the microstructural changes in the thin porous layers (Table 2). However, in the present case, the measured bulk density values could not be directly correlated with the apparent porosity values of the entire system due to its laminar structure. Average density values were calculated using the measured bulk density values and the theoretical density value of thermally stable α-alumina.

### 3.2 Dynamic loading of dense alumina

Typical raw signals measured by strain gauges mounted in the middle of the incident and transmitted bars are shown in Fig. 6(a) for the case of dense alumina disks. The variations of voltage signal with a common starting time scale of the incident, reflected and transmitted signals are given in Fig. 6(b). Fig. 6(c) shows typical input and output forces calculated from the data shown in Fig. 6(b) using Eqs. (1) and (2). The good overlap between the input and output pulses indicates that an equilibrated state of the specimens was achieved during the dynamic impact experiments. Since dynamic equilibrium was achieved during the experiments, hereafter only the output force pulses were compared.

In general, dense ceramic specimens exhibit axial cracks parallel to the loading axis once the applied force reaches the threshold limit under compression[4]. However, in the case of dense alumina specimens, only one case out of several samples tested at these rates and load levels exhibited catastrophic failure (dashed line in Fig. 7(a)). Even though the same precursors or green tapes were used to fabricate all the dense alumina studied, failure of a sample might be related to defects present in the sample which is a very common and generally unavoidable problem associated with ceramic processing. Fracture in the sample during dynamic loading can be detected by a sudden drop in the transmitted force before the end of the loading pulse, in comparison with the gradual decrease of force at the end of loading in the intact specimens (Fig. 7(a)). Fig. 7(b) shows only the loading output curve of the alumina disk that failed (same as dashed line in Fig. 7(a)). The average diametrical compression rate which was calculated through Eq. 3 was about 0.63 – 1.0 m/s and the values have been marked in Fig. 7 and subsequent figures as an indicator of the severity of the impact event in each case. High speed images of the failure processes corresponding to the points highlighted in Fig. 7(b) are shown in Fig. 8. The first image in Fig. 8, denoted 0 $\mu$s, shows the dense alumina specimen compressed between two rods in the testing section of the SHPB just as the loading wave arrives on the left hand side of the sample. The second image corresponding to 53 µs, shows the first instant of visible crack initiation. At this time, nearly 80% of the applied force was acting on the sample. The cracks initiated from the contact area and propagated in an elliptical direction, indicating that the material layering does not play a role in the failure path, eventually fracturing the entire sample. In the third image of Fig. 8, cracks have propagated further along the axis, and still the peak force had not been reached. By the time the crack propagated to the other end of the specimen at around 66 µs, the overall compressive force reached its maximum value. The sequential high-speed camera images revealed that substantial cracking occurred well before the threshold failure force. The fifth image, Fig. 8(v), was taken 6.67 µs later, and major cracking is visible on the surface. Fracture through several cracks has destroyed the specimen into several fragments which are visible from the sixth image onwards, corresponding to the unloading part of the experiment.

The overall force *vs.* time response and failure process of the dense alumina specimens was in agreement with the reported compressive behavior of alumina which indicated the initiation of a crack at an early stage[4,6,13]. The compressive behavior of dense alumina specimens indicated that the applied force was transmitted from incident to transmitted bar and the amount of transmittance depended on the impact loading (Fig. 7(a)). The overall transmitted force varied, depending on the amount of pressure on the incident striker, which depended on the pressure on the gas chamber. Unless dissipated through the fracture most of the applied impact energy on the specimen was transmitted. In the case of intact samples, no major or visible macrocracks were observed after the dynamic loading event. However, microscopic analyses revealed that microcracks initiated at the point of contact, possibly due to the localized shear deformation, as shown in Fig. 9. In Fig. 9(a) microcracks are shown around the contact point and they propagated in the same elliptical direction in which other samples catastrophically failed, perhaps indicating incipient failure conditions. The tip of the microcracks in the intact specimens appeared blunt when compared to the microcracks on other fractured specimens. The magnified view of the microcracked region indicated that the removal of material within the cracked region was by an intergranular fracture mode, Fig. 9(b). As expected, the microstructure of catastrophically fractured specimens, which corresponded to Fig. 7 (b), also followed the intergranular fracture (Fig. 9(c)), since the applied force as well as the strain rate were not enough to cause transgranular fracture or twinning[4,37]. Fracture steps were associated with most of the surface, indicating the fracture path on the grain boundaries, which can clearly be seen (Fig. 9(b) and (c)).

### 3.3 Dynamic mechanical response of laminates

In the case of alumina/porous-alumina layered systems, disks were aligned in parallel (0° orientation) and perpendicular (90° orientation) to the layers (inset of Fig. 2). The range of samples impacted in two different orientations clearly revealed two different mechanical responses. Most of the impact force was transmitted in the case of the 90° orientation, whereas it was observed to undergo modest to severe fracture at 0° orientation (Fig. 10). As in the dense alumina specimens, layered ceramic structures also exhibited a force *vs.* time response depending on the impact velocity in both the orientations.

The repeatability of the dynamic experiments on brittle materials such as ceramic is critical. Figure 10 shows two replicates of the extreme cases of transmitted force signals of specimens which have different microstructure (20% and 50%) impacted under two different loading angles (0° and 90°). It can be seen that the repeatability of the experiments on the specimens under 90° loading angle was better compared to the specimens under 0° loading angle which shows some difference, especially for 20 vol % (Fig. 10(a)). However, all the specimens impacted at 0° catastrophically failed during the experiments. Since the crack propagation in such brittle, nonhomogeneous materials (in the case of laminates) depends on various factors and consequently may show very different behavior once crack initiates, we decided to focus on the force vs. time behavior before crack initiates. These crack initiate points have been noted in Fig. 10. Note that both for 20% and 50 % under 0° loading the response is nearly identical before cracking initiates. For the 50% case, the first kinks on the signals, which indicate the crack initiation, are almost identical also. These results indicate that this experimental configuration produces repeatable results until failure initiates. In addition, high speed images are shown in Fig. 11 for two experiments with 20 vol% and two experiments with 50 vol%, both under 0° loading. Specimens for two repeated experiments exhibit very similar, though of course not identical, failure patterns. Therefore even post-failure although the transmitted force is not as repeatable as in the pre-failure regime, the failure modes are quite repeatable – certainly enough to differentiate between differences in different configurations (see Fig. 14 subsequently). The average diametrical compression rate which was calculated through Eq. 3 was about 0.41 - 1.14 m/s and the values have been marked in Fig. 10.

### 3.3.1 20 vol% in 0° orientation

The compressive force *vs.* time response of the sample containing 20 vol % porosity in the interlayers is shown in Fig. 10(a). The dynamic failure process of the laminates in 0° orientation is shown in Fig. 11. In 20 vol % samples, the force *vs.* time curves followed the same trend until the specimen impacted in the 0° orientation started to deviate due to crack initiation at around 46 $\mu$s (Fig. 10). The failure process was very similar to the fractured alumina specimen discussed previously. In the 20 vol % specimen, axial cracks were initiated and propagated in the axial direction as seen in the images taken at 6.67 $\mu$s intervals (top row in Fig. 11). The crack initiated when the overall compressive force reached ~ 60% of its eventual peak force. The cracks grew as the loading increased, as shown in the images. The layers through the contact axis underwent severe fracture and the applied force decreased by dissipating its energy through delamination and crack deflection. The last two frames in Fig. 11 (for 20 - 0° - a), show that the specimen could no longer sustain the load, and resulted in massive failure through the layers.

The fracture analysis of the 20 vol % in 0° orientation is shown in Fig. 12(a). After the impact, the specimens appear catastrophically damaged, as shown by the first and second images of Fig. 12(a). Once the cracks were initiated during impact, they eventually grew and propagated along the compressive loading axis, mainly through the weak layers or at the interface between the weak and strong layers. The crack network led to debonding and disintegration of the laminates. The microscopic analysis also revealed crack branching and crack deflection at the interfaces (second image of Fig. 12(a)). In some locations, cracks were deflected and remained within the porous interlayer itself, as shown in the bottom arrow marked region of the third image in Fig. 12(a). At the same time, it is important to note that if any of the cracks nucleated within the strong layer itself in the axial direction, they did not always need to be deflected to the weak layers but could also propagate in the axial direction itself, as shown in the fourth image of Fig. 12(a).

### 3.3.2  20 vol% in 90° orientation

The transmitted force, measured in the case of the 90° orientation, resembled the trend observed on the intact dense alumina specimens (Fig. 10 (a)). In the 90° orientation, it was hard to observe fragmentation in the case of 20 vol % samples. Even after complete unloading of the impact force, the specimen appeared intact in the high speed images (first image of Fig. 12(b)). However, in the localized regions, small portions of the specimen chipped away near the transmitting bar, at a much later stage (second image of Fig. 12(b)). This cracking phenomena at the later stage might be due to the relaxation of the applied compressive force or tensile waves generated after the compression impact. This behavior was similar to the well-studied chipping behavior of brittle solids (both ceramics and glass) under uniaxial/sharp contact loading conditions[38,39]. The absence of severe fragmentation or a severe fracture event which would help to dissipate the impact energy resulted in microcracking of the region which was far beyond the impact zone (third image of Fig. 12(b)). Further microscopic analysis around the impacted region revealed crack branching and crack deflection at the interfaces (fourth image of Fig. 12(b)). The microscopic observation of the fractured fragments, fifth image in Fig. 12(b), also showed fracture steps caused by intergranular fracture as well as the influence of porosity on the crack path. However the axial cracks did not coalesce readily, since this would involve crack propagation vertically which is not favorable in this case, and therefore the sample never completely fragmented.

### 3.3.3  50 vol% in 0° orientation

The crack pattern obtained for 50 vol % specimens at the 0° orientation resembled those observed in the 20 vol % sample except that minor fragmentation took place in severely affected porous layers, Fig 13(a). Several microcracks formed and propagate either at the interfaces or in the layers (both strong and weak), before dominant cracks could propagated through the specimens. In order to clearly reveal the fracture process which occurred during the dynamic failure event, the impacted zone was observed under the same magnification but at different conditions (first and second images in Fig. 13(a)). The circled region in the second image clearly shows the missing fragments in the weak layers due to the impact. However, it was hard to observe in directly impacted regions (since they could not be recovered) and it could be observed only next to the layers which were shattered during the failure event.

### 3.3.4  50 vol% in 90° orientation

The fracture analysis on the 50 vol % specimens at the 90° orientation, however, revealed an interesting failure mechanism compared to 20 vol % at the 90° orientation (Fig. 13(b)). Even in this case, chipping of some of the specimen around the region in contact with the transmitted bar took place at a much later stage, and fragmentation appeared severe around the impact region. The first and second images in Fig. 13(b) revealed that the fragmentation in the weak layers took place only nearer to the impacted zone along with crack deflection. Even in the 90° orientation, the major failure resulted in the breaking of some small parts of the specimen which occurred predominantly along the interface between the weak and strong layers.

In general, dynamic failure processes were similar for the 20 and 50 vol % laminates loaded in the 0° orientation with severe fracture around the impacted zones. However, microscopic analysis revealed that the 20 vol % specimen was more extensively damaged in the axial direction than was the 50 vol %. Notably, 20 vol % samples held the impact force for a shorter time in total than did the 50 vol % samples (Fig. 10). Furthermore, the 20 vol % samples eventually failed in a shorter time between the crack initiation regime and complete failure. In other words, the 50 vol % specimens were able to bear further increased axial force but cracked much earlier (Fig. 10). In both 20 and 50 vol % samples, even though crack deflection took place between the strong to weak layers and within the weak layer itself, major events which contributed to fracturing the specimens into several segments primarily occurred at the interface between the porous and strong layers (Fig. 11). In 20 vol % samples, when extensive cracking took place in the 0° orientation, no sign of microcracks was observed on the regions which were far beyond the impacted region. Previous studies on granite specimens under various strain rates showed that the crack initiated at the center of the specimen and split into two halves and also applied force which shear deformed the specimen around the contact zone[40]. However, ceramic laminates in the present study behaved like metallic-intermetallic laminates in 0° orientation, where cracks initiated near the contact points and grew along the loading axis[23].

In the 90° orientation, almost none of the studied samples showed any sign of cracking during the experiment (load/unloading) time. However, the area around the contact zone tore off at a much later stage (tensile splitting). When the disks were loaded perpendicular to the layers (in the 90° orientation), the influence of dense alumina layers could have changed the fracture process significantly in both the 20 and 50 vol % laminates. The overall fracture behavior of laminates having two different microstructures, namely the 20 and 50 vol % graphite added porous interlayers, behaved in a similar way when they were impacted at 0° and 90° orientations (Fig. 10 and 11). Such a striking difference in mechanical responses under 0° and 90° orientations implied that the failure mechanisms were significantly different.

### 3.4 Energy dissipation mechanism

The real-time high-speed images and post mortem microscopy analysis of the failure process indicates that fracture in these materials was associated with features over several length scales, predominantly with microscopic features such as pores. Researchers in the dynamic fracture behavior for alumina-porous alumina laminates reported that for a fixed porosity level in the weak layers, the yield stress and strain increased with increasing impact velocity and strain rate[32]. It has been widely reported for various materials that the compressive strength increases with increasing applied strain rate due to the associated micro-mechanisms (microcracking, dislocation activity, and phase transformation)[13,23,40,41]. Even in the case of static analysis for the similar systems, reports showed that increasing porosity promoted crack deflection and hence increased fracture toughness until the porosity values reached a threshold limit[31]. In the present study, increasing the porosity in the weak layers from 20 to 50 vol % in 0° orientation increased the holding time, and could even sustain large impact and dissipate energy in the 90° orientation (Fig. 10 and 14). This behavior could have been caused by the following two reasons: (i) Collapse of porous matter under compression (cushioning effect) which helps load redistribution as reported in several systems under various strain rates[31,32,36]. (ii) Fragmentation due to a more refined microstructure in the weak interlayers. In our case, the microstructure of the porous interlayers in the case of the 50 vol % underwent severe refinement due to the addition of graphite particles (Table 2). Hence, the loosely packed particles in the porous layers might have acted as an effective toughening mechanism by fragmentation. Moreover, the volume fraction of the porosity in the layered systems decides the relative fracture energy of the adjacent layers which could have played a role in the crack deflection mechanism[42,43]. It can be seen from Figs. 12 and 13 that porous interlayers were effectively deflecting the cracks which provided toughening to the system. As shown by Ma et al.,[31] on similar systems, interactions between the homogeneously distributed porosity led to fracture of the alumina layer in the porous interlayers which caused cracks to kink out of the interface[20,44]. Finally, a significant fact should be noted that on the basis of our evaluation, not only the orientation of porous alumina layers but also the volume of porosity in the weak layers impacted the overall force absorption or transmission.

The present investigation shows that by controlling microstructure and orientation of impact, stress can be transmitted without causing damage or dissipated within the system itself (Fig. 14). The intact samples did not mean that they were unbreakable. By applying a higher load or higher strain rate or combination of both, they will eventually shatter. Nevertheless, what is most important to note in the present investigation is that layering the material is an effective way to mitigate the applied force (or stress) along a desired direction. In any given case, by using the same microstructure of the lamellar structure, stress can be transmitted or dissipated, based on the requirements. By tailoring the microstructure in the weak layers with more refined pore and grain sizes (e.g., 50 vol % in the present study), fragmentation in reduced amounts can help to limit microcracking in the regions which are far away from the impact zone.

## 4. Conclusions

Dense alumina as well as alumina/porous-alumina laminates were successfully fabricated in dimensions suitable for a Brazilian disk experiment. The experimental technique and sample specification followed in this study helped us to investigate the role of microstructure and orientation of impact on stress mitigation. The applied force transmitted to the receiving end of a SHPB set up highly depended on the microstructure and the orientation in which the samples were impacted. In the layered systems, most of the impact energy was dissipated through severe fracture in 0° orientation. Cracking began near the incident bar and grew in an axial direction towards the transmitted bar, producing axial splitting as the failure mode. Stress-wave mitigation through channeling cracks into interlocking configurations enhanced energy dissipation through fracture. However, the failure mode varied significantly and most of the applied force was transmitted in the case of the 90° orientation. Fragmentation and crumbling of porous interlayers could have provided load redistribution in the case of 90° orientation and might have acted as an effective energy absorber. The severity of damage around, as well as far beyond the impact region appeared to be less in 50 vol % specimens compared to 20 vol % which might be related to a more refined microstructure in the former case.

**Acknowledgments**

Derrick N. Poe is acknowledged for his help with the initial stage of sample fabrication. The United States Army Research Office MURI grant (W911NF-09-1-0436), through Dr David Stepp, is acknowledged for funding this work. CJES is indebted for a National Science Foundation Graduate Research Fellowship Program (NSF-GRFP, grant no. NSF DGE 11-44245). The SEM work was carried out in the Frederick Seitz Materials Research Laboratory at the University of Illinois at Urbana-Champaign, which are partially supported by the U.S. Department of Energy under grants DE-FG02-07-ER46453 and DE-FG02-07-ER46471.

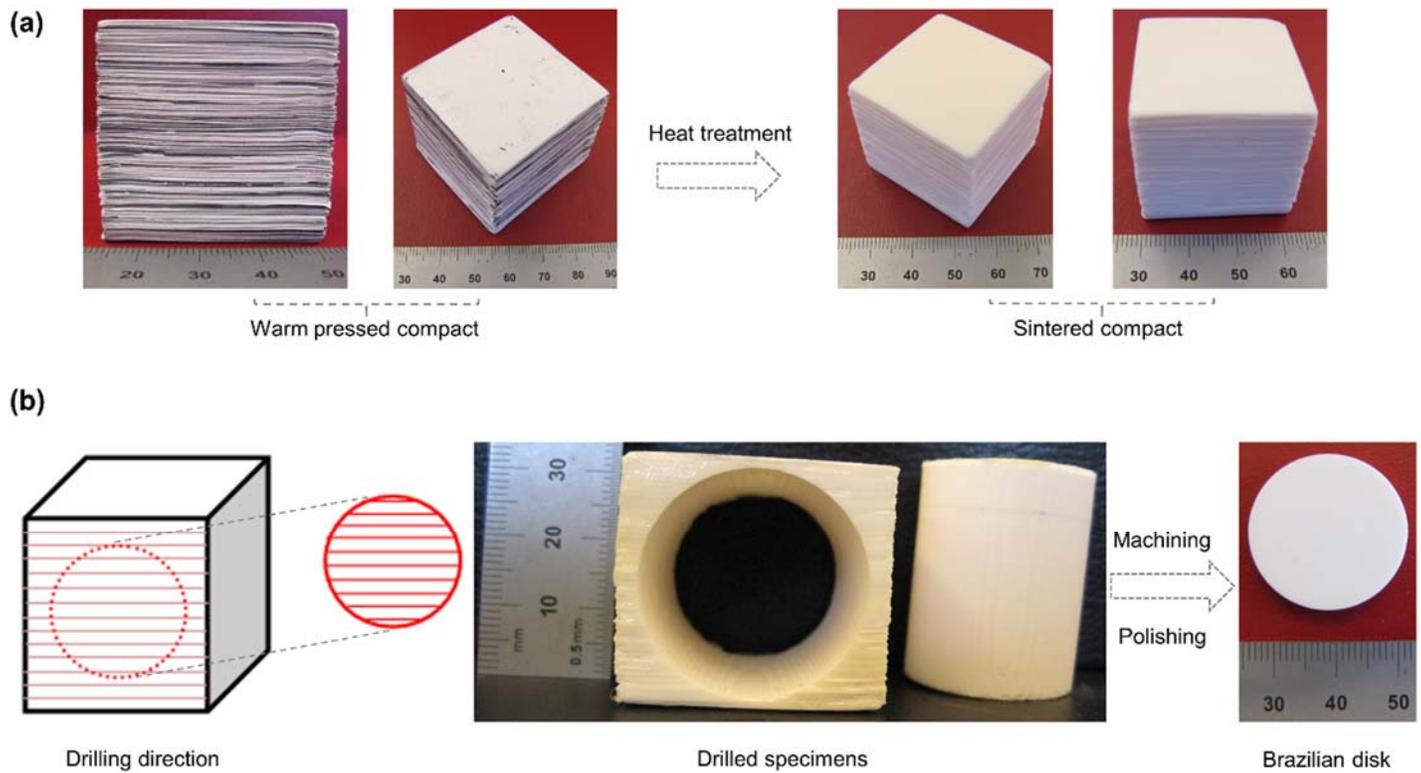

**Fig. 1** Illustration of Brazilian disk fabrication procedure. (a) Photographs of the warm pressed compacts consisting of alumina and graphite-containing alumina green tapes, and the sintered compact. (b) Schematic representation of the direction in which cylinders were drilled out, photographs of the drilled cylinder, and the Brazilian disk sliced out of the cylinder.

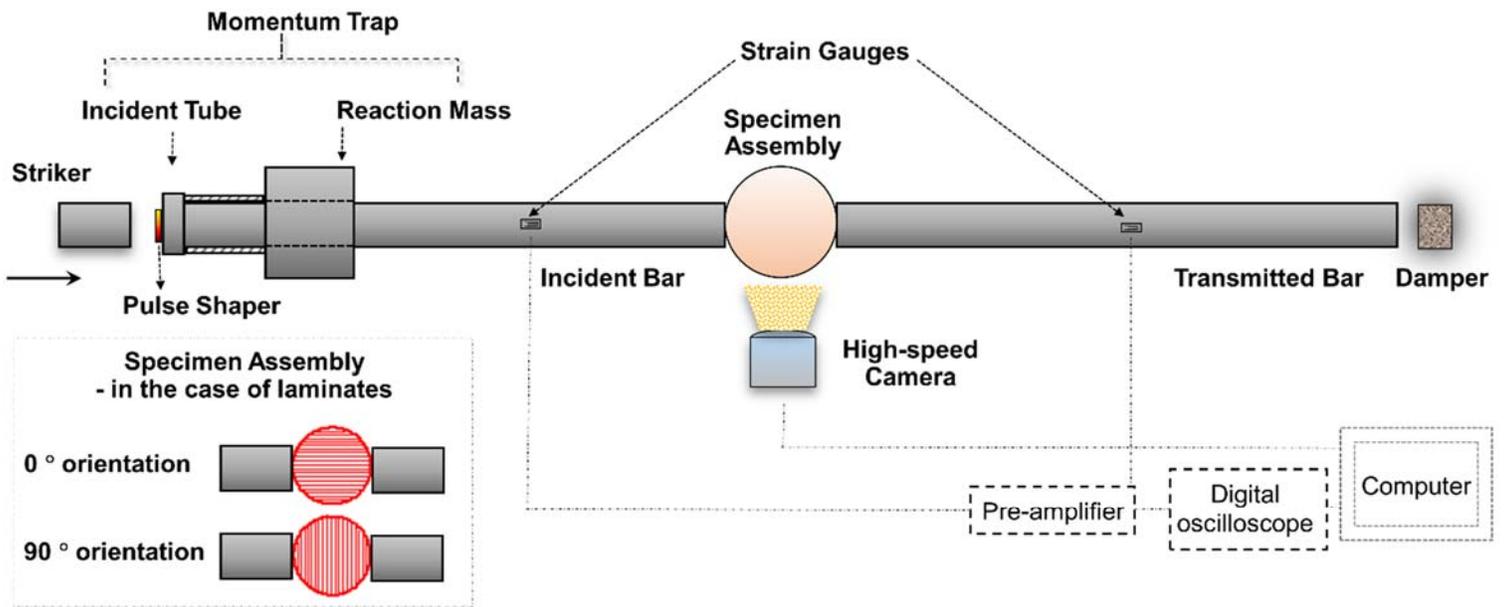

**Fig. 2** Schematic of SHPB apparatus with momentum trap and pulse shaping as used in this study. The inset illustrates the orientation of laminated samples placed between incident and transmitter bar.

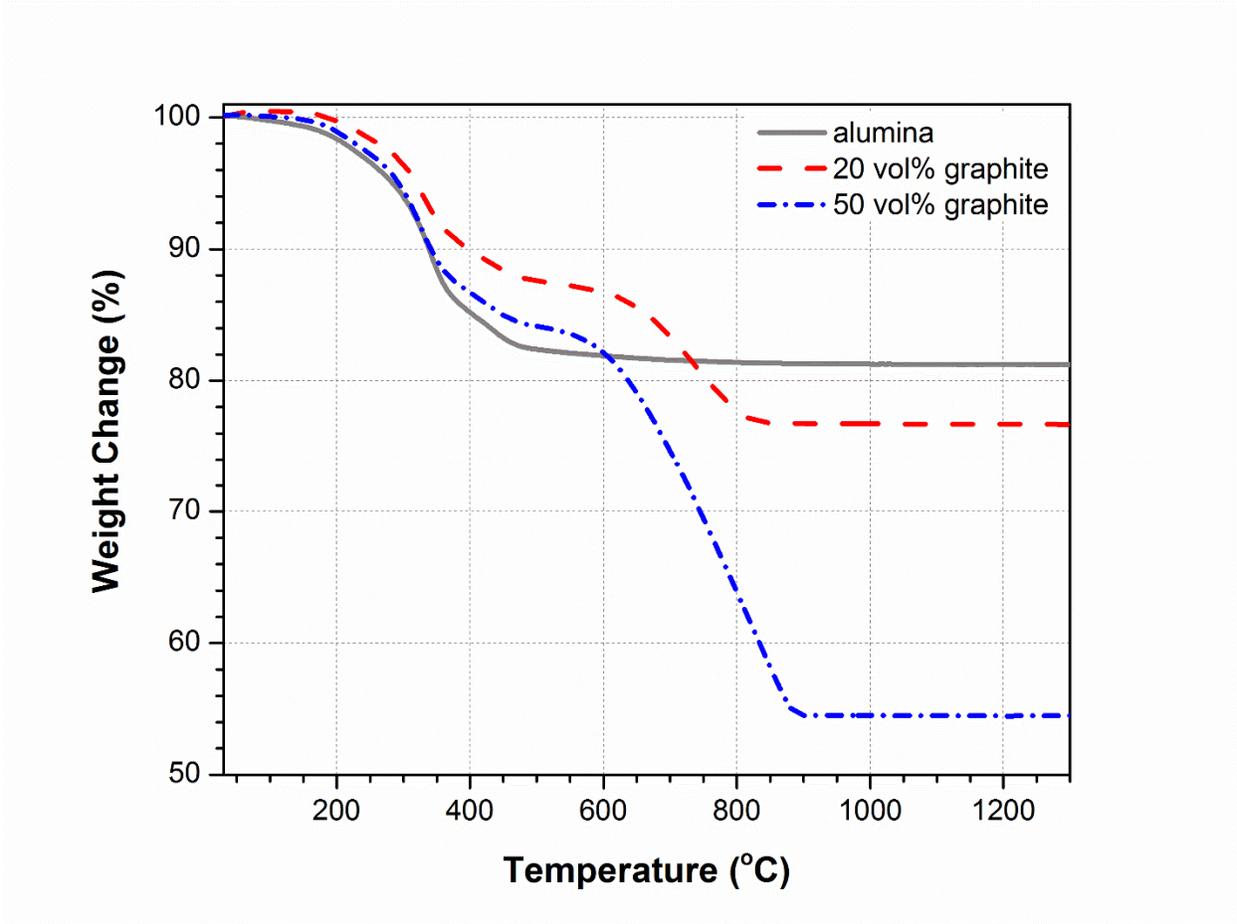

**Fig. 3** TGA of the as-prepared green tapes containing 0, 20 and 50 vol % graphite particles.

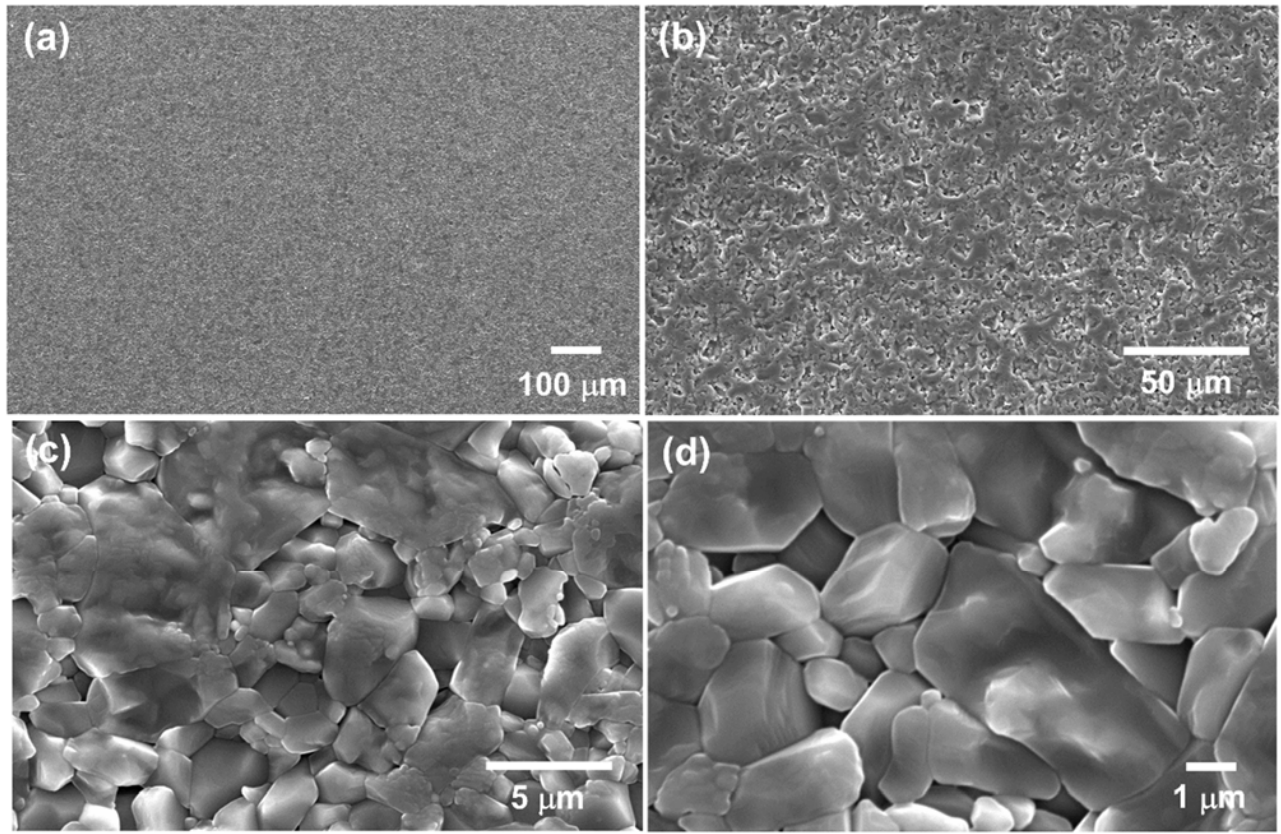

**Fig. 4** Microstructure of the sintered alumina at 1550 °C for 5 h. (a) and (b) as polished surface at two different magnifications. (c) and (d) thermally etched at 1400°C for 1 h to reveal the grain boundaries.

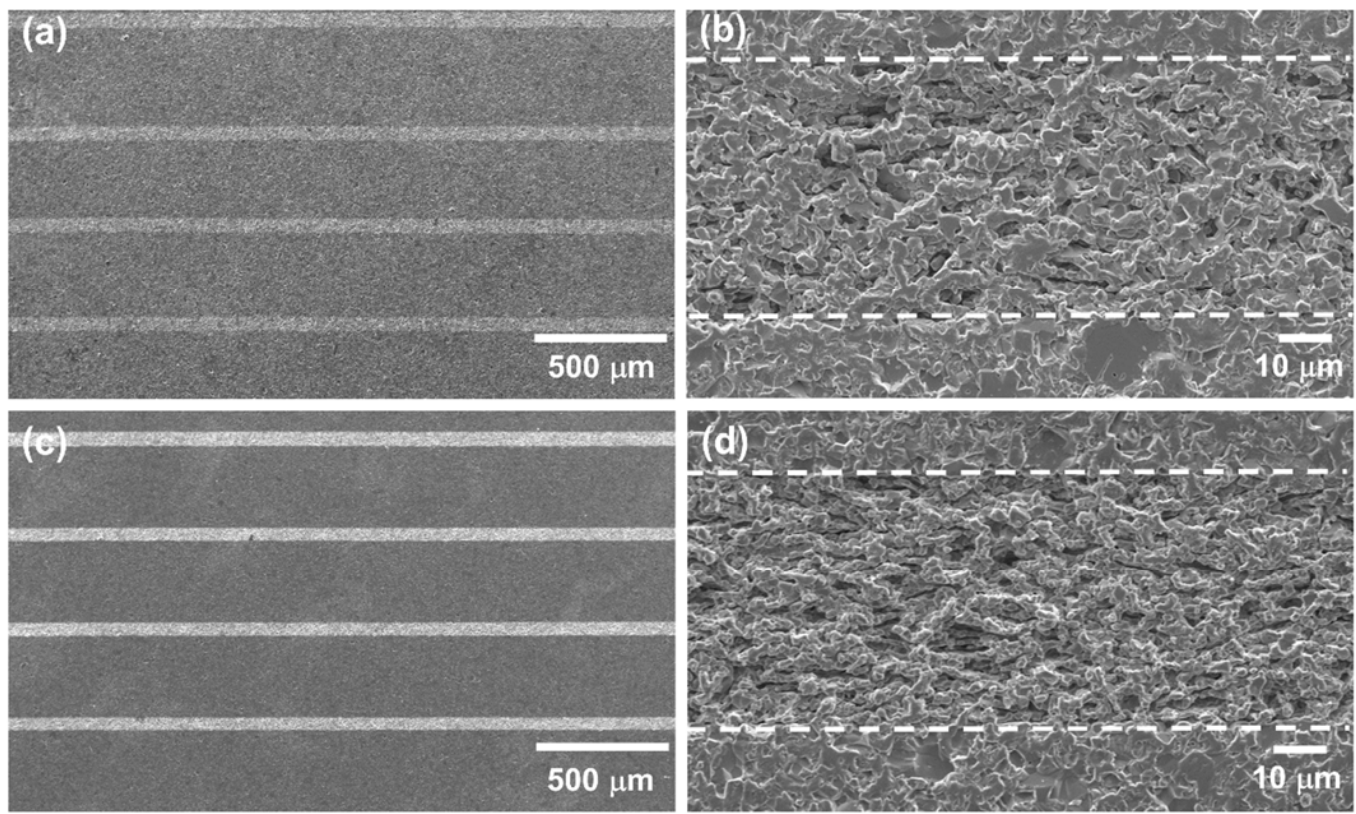

**Fig. 5** Microstructure of the sintered laminates containing 20 vol % (a) and 50 vol % graphite in the porous layers (c), magnified porous regions of (a) and (c) are shown in (b) and (d), respectively. (The dotted lines in (b) and (d) are a guide for the eyes).

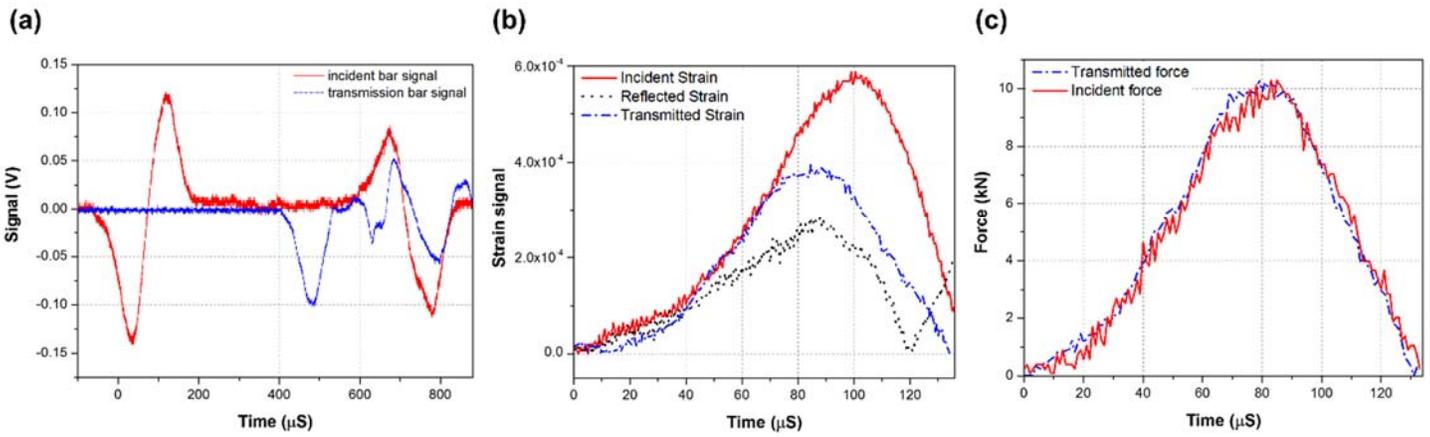

**Fig. 6** (a) Example of typical raw data recorded from momentum trapped bar, (b) computed strain value, and (c) incident and transmitted force *vs* time (results shown are on intact dense alumina shown in Fig. 7(a)).

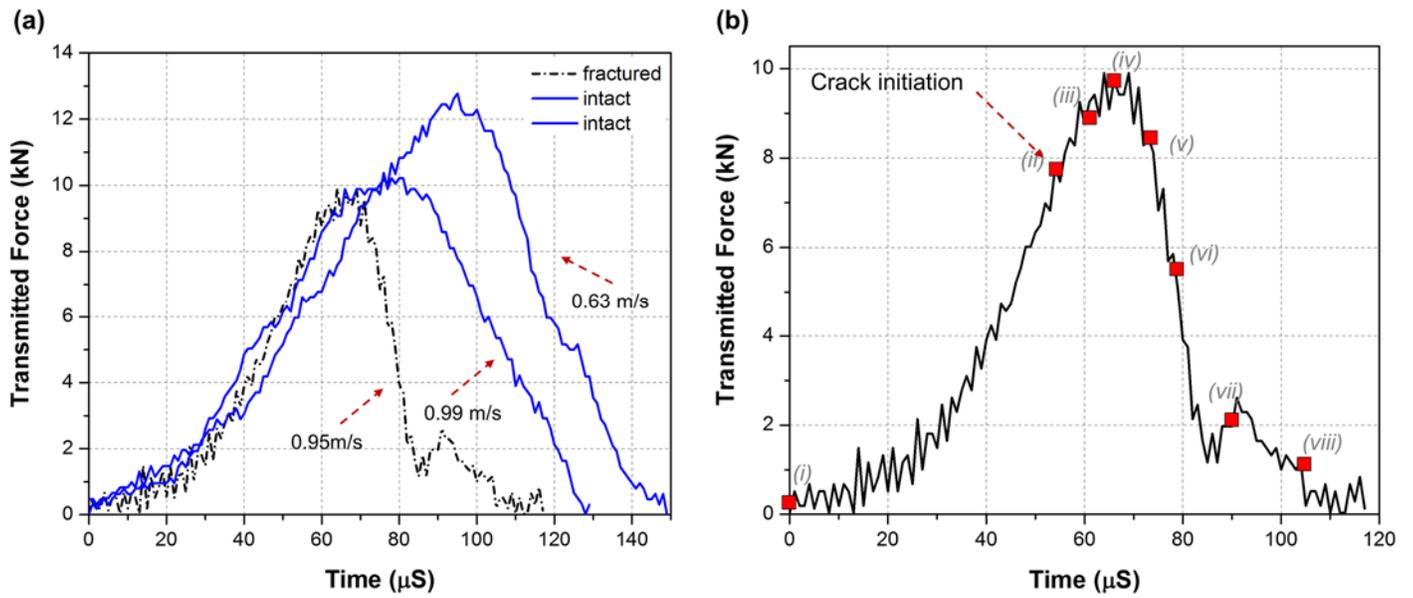

**Fig. 7** Transmitted force of studied dense alumina specimens in (a) and (b) is the fractured specimen shown in (a). The diametrical deformation rate for the Brazilian disk specimens calculated using Eq. 3 is shown in (a).

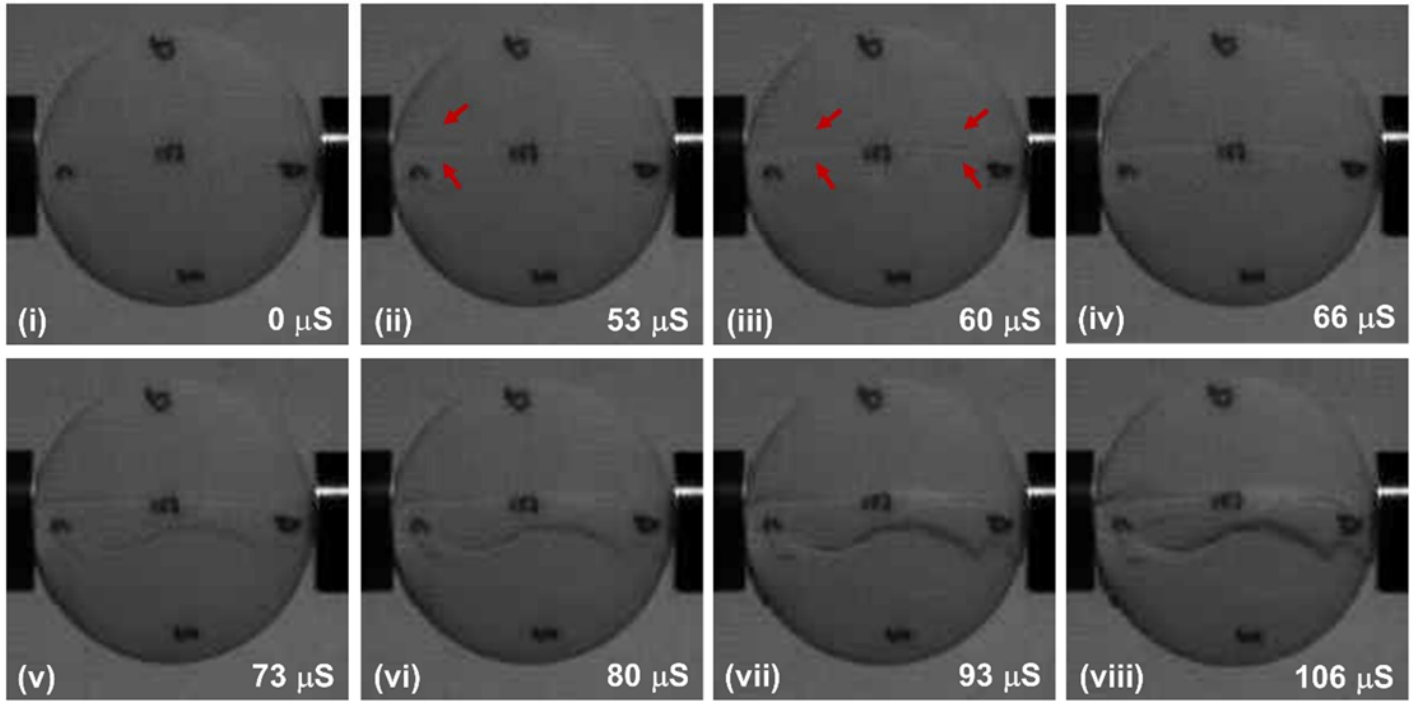

**Fig. 8** A selected sequence of high-speed camera images showing dynamic failure and fracture process of alumina illustrated in Fig. 7(b). (Here the incident bar is on the left and the transmission bar on the right side of the sample). The crack initiation process is highlighted using red arrows. The time corresponds to each of the photographs is shown on the force-time curve in Fig. 7(b).

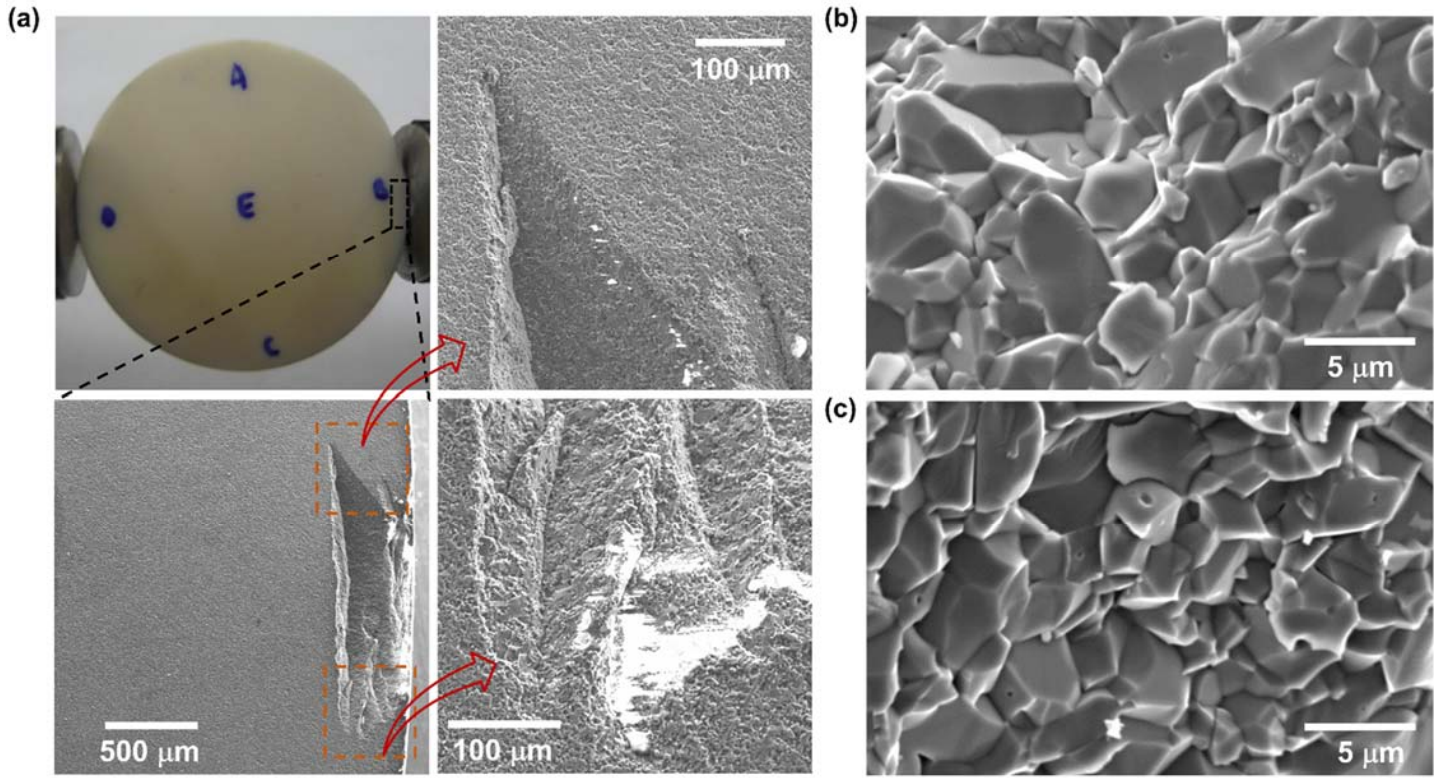

**Fig. 9** (a) Photographs and SEM images describe the crack initiation regime in the intact alumina specimen shown in Fig. 6 and Fig 7(a). (b) Magnified image of the crack initiated region of (a), and (c) fractured region of Fig. 8.

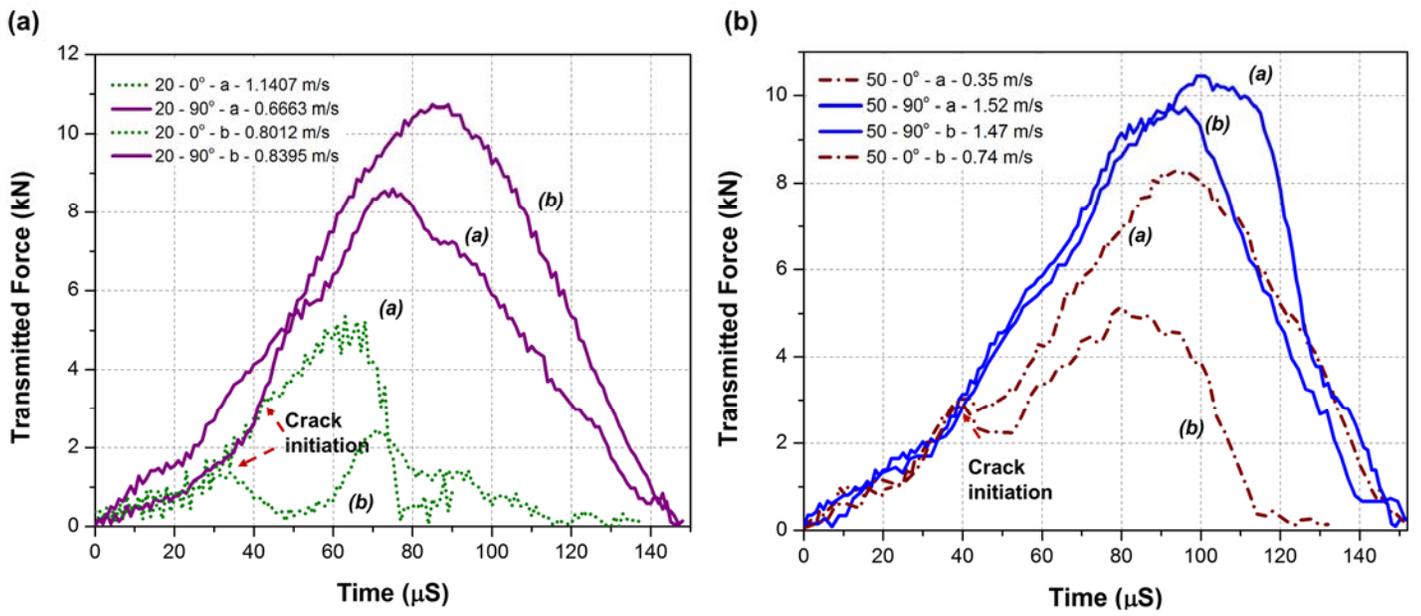

**Fig. 10** (a) Transmitted force *vs* time corresponds to 20 vol % sample at 0 and 90° orientation and (b) for 50 vol % samples.

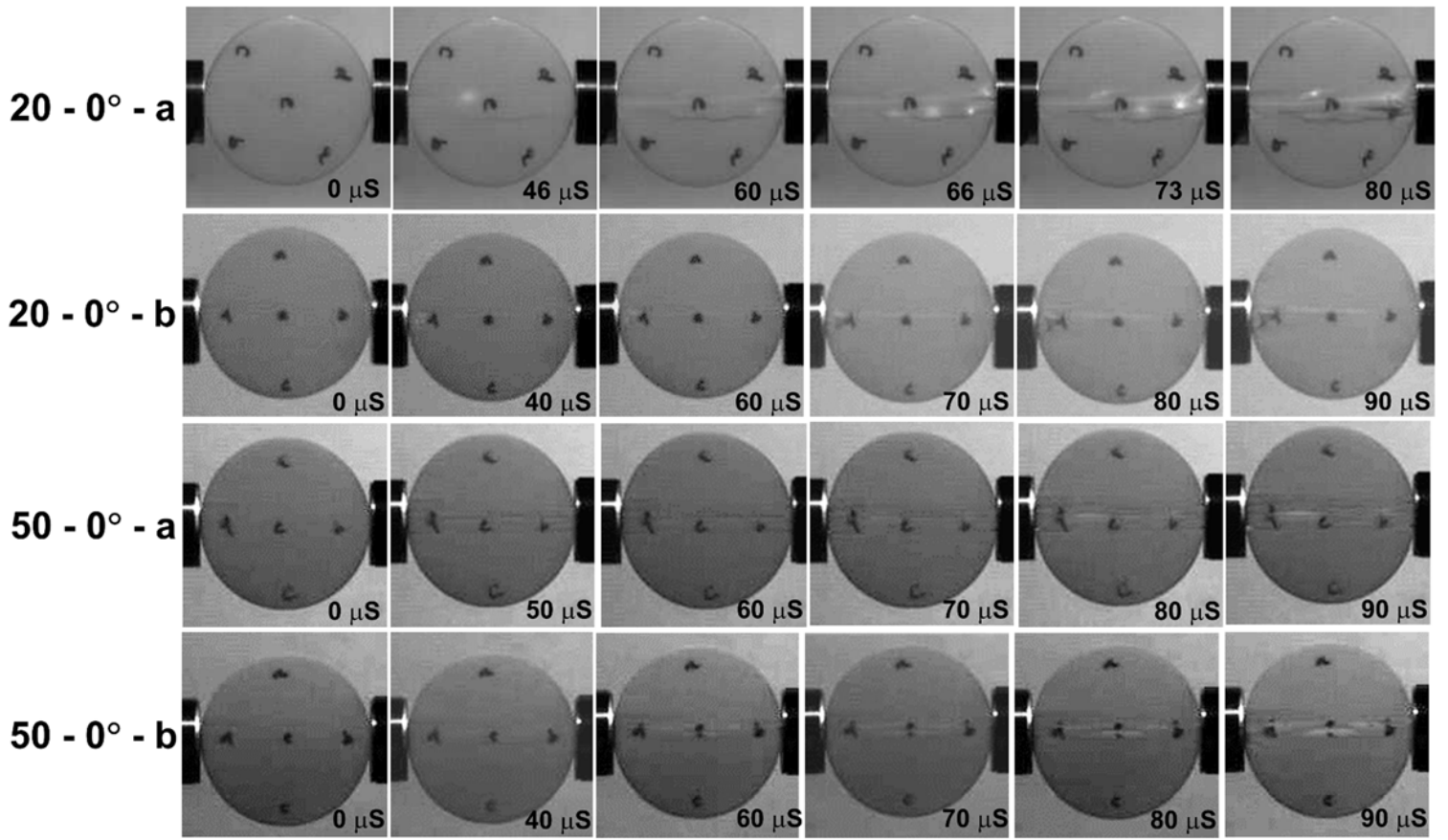

**Fig. 11** Dynamic failure and fracture process corresponds to 0° orientation in Fig. 10(a) and 10(b). The time corresponds at which each of the photograph taken are shown on photographs.

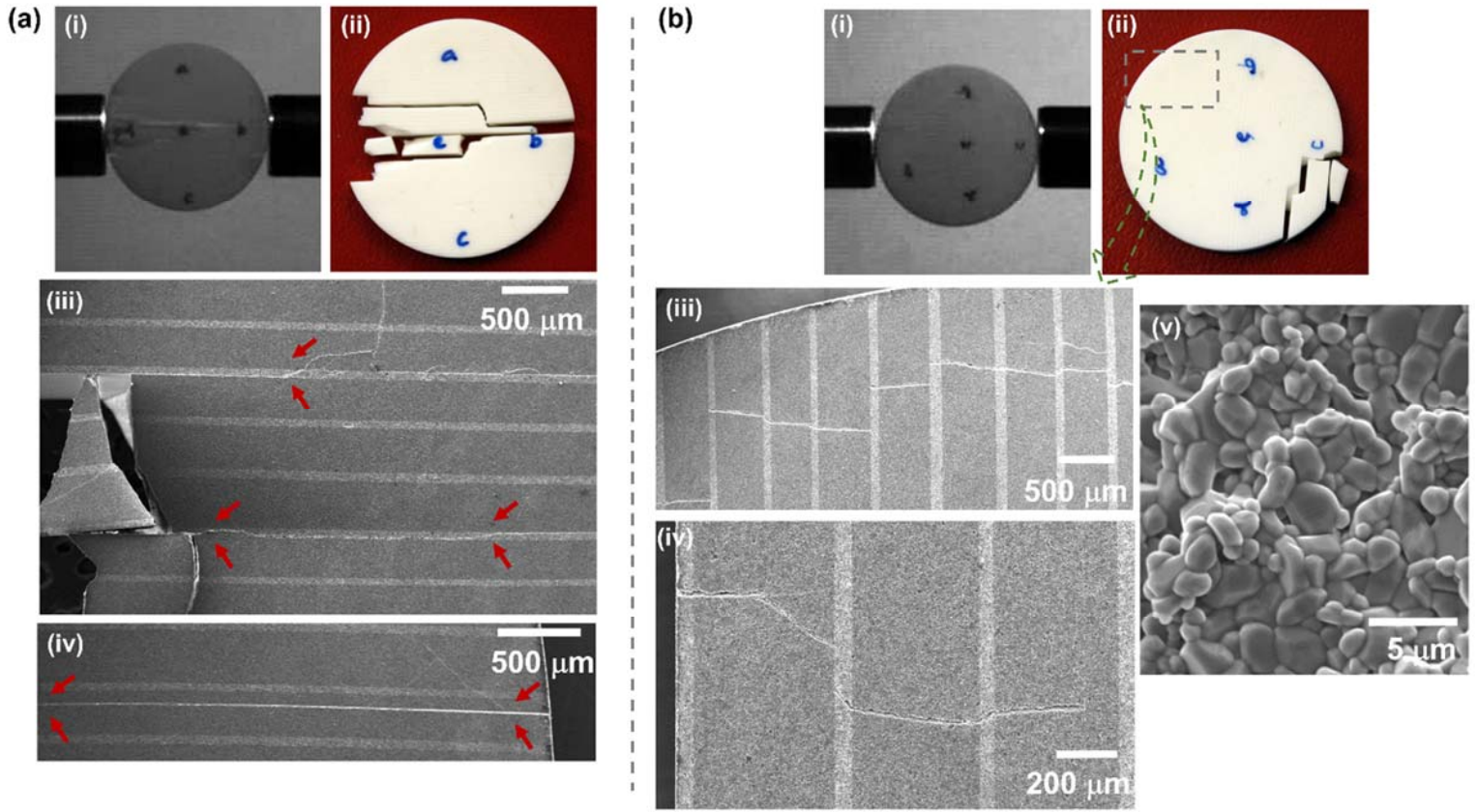

**Fig. 12** The fracture and failure process associated with 20 vol % sample at (a) 0 and (b) 90° orientation. In (a), (i) sample at the end of the experiment, (ii) specimen recovered after the experiment, (iii) fractured images show crack propagation and deflections within the porous layers, (iv) crack propagation in the dense layer as well. In (b), (i) sample at the end of the experiment, (ii) specimen recovered after the experiment, (iii) microcracks on the specimen which was far away from the compressive zone, (iv) crack deflection, (v) and (vi) fractured wavy surfaces with intergranular fracture.

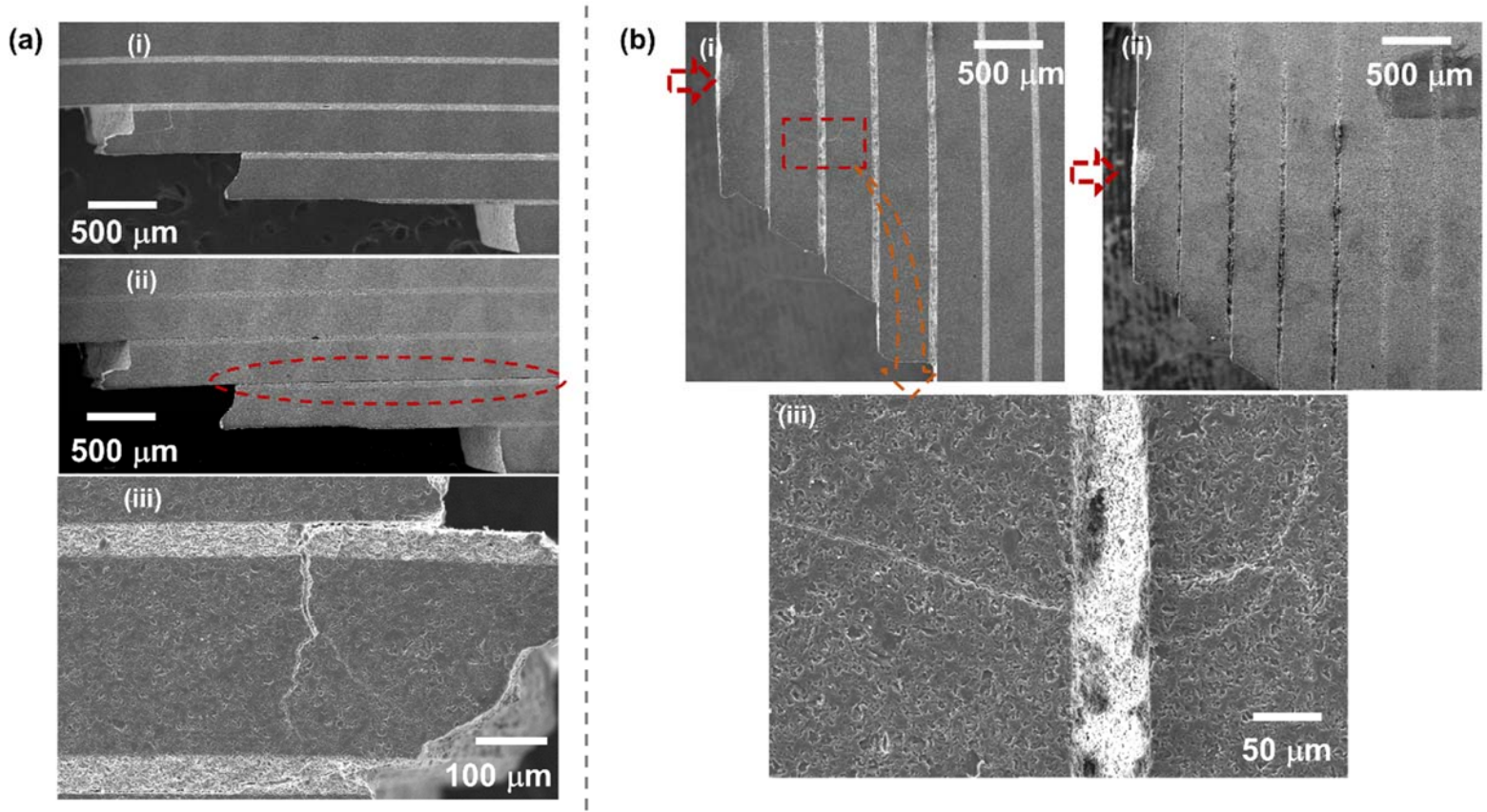

**Fig. 13** The fracture and failure process associated with 50 vol % sample at (a) 0 and (b) 90° orientation. In (a), (i) crack deflections and propagation on the fractured surface, (ii) same as (i) but observed under different conditions to show the fragmentation (circled region) near the highly damaged region, (iii) crack deflection. In (b), (i) and (ii) are at the same location but under different conditions to reveal the fragmentation on the weak layers around the impact zone, (iii) crack deflection around the impacted zone. The big red color arrow in (i) and (ii) indicates the contact zone.

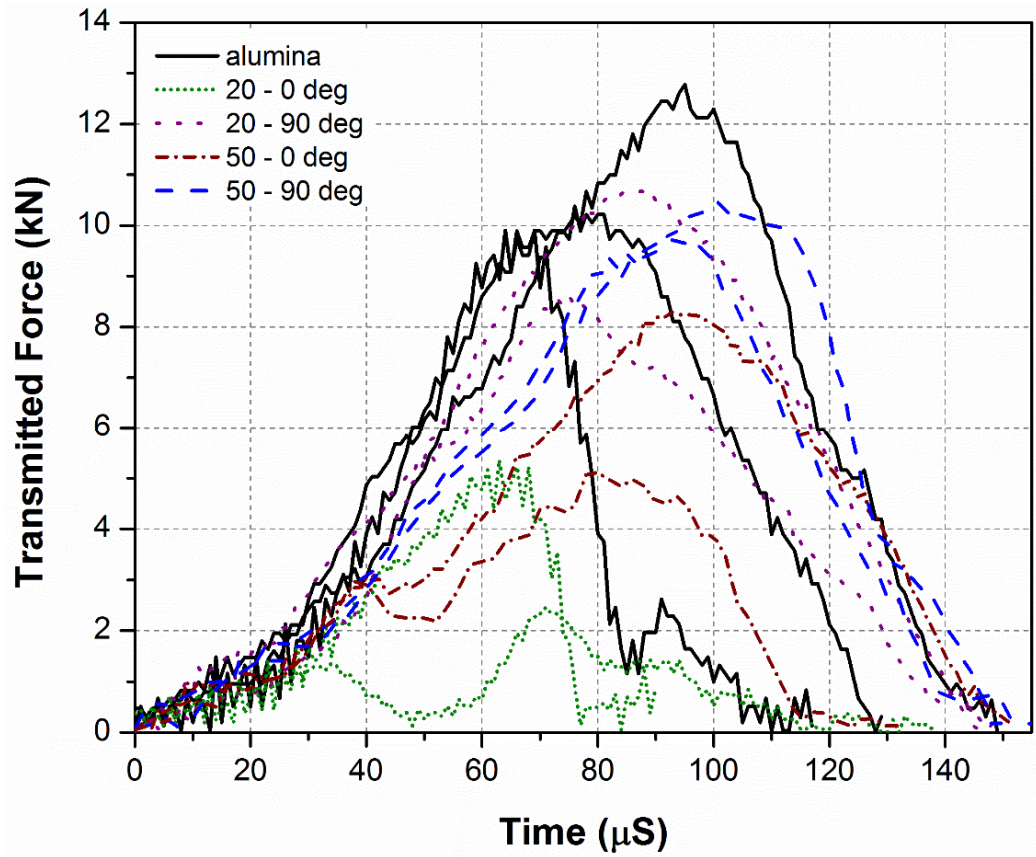

**Fig. 14** Evolution of transmitted force-time response for alumina and alumina/porous laminate in different orientations.

**Table 1.** Formulation used to fabricate alumina tapes with and without graphite particles.

| Sample ID | 1st stage milling (for 16 h) | | | | | | | 2nd stage milling (for 4 h) | |
|---|---|---|---|---|---|---|---|---|---|
| | Composition | Powder (%) | DI water (%) | Binder (%) | Plasticizer (%) | Defoamer (%) | | Binder (%) | Defoamer (%) |
| Dense alumina | Alumina | 53.4 | 21.8 | 11.85 | 0.5 | 0.3 | | 11.85 | 0.3 |
| 20 vol % | Alumina-20 vol % graphite | 55.8 | 25 | 9 | 1 | 0.1 | | 9 | 0.1 |
| 50 vol % | Alumina-50 vol % graphite | 51 | 27 | 11.5 | 1 | 0.3 | | 9 | 0.2 |

**Table 2.** Physical properties of the dense and laminated samples.

| Sample ID | Bulk density (g. cm$^{-3}$) | Average density (% TD) | Grain size ($\mu$m) (Range) | Pore size ($\mu$m) (Range) |
|---|---|---|---|---|
| Alumina | 3.93 ± 0.01 | ~ 98.5 | 2.6 (0.5 – 13.2) | 0.65 (0.2 – 1.6) |
| 20 vol % | 3.84 ± 0.01 | - | 1.9 (0.45 – 5.3) | 4.6 (0.2 – 16.3) |
| 50 vol % | 3.76 ± 0.01 | - | 1.3 (0.44 – 3.5) | 3.3 (0.1 – 18.6) |